\documentclass[twocolumn,floatfix,showpacs,pra,aps,a4paper]{revtex4}

 \usepackage{bm}
 \usepackage{amsmath}
 \usepackage{amssymb}
 \usepackage{latexsym}
 \usepackage{amsfonts}
 \usepackage{epsfig}
 \usepackage{subfigure}
 \usepackage{color}
 \definecolor{darkblue}{rgb}{0,0,.5}
 \usepackage[linktocpage, colorlinks=true ,linkcolor=darkblue, citecolor=darkblue]{hyperref}
 \usepackage[all]{hypcap}

\newcommand{\ket}[1]{\left|#1\right>}
\newcommand{\bra}[1]{\left<#1\right|}
 
 \newcommand{\nn}{\nonumber\\}

 \newcommand{\bea}{\begin{eqnarray}}
 \newcommand{\ea}{\end{eqnarray}}
 \newcommand{\eea}{\end{eqnarray}}
 
 \newcommand{\trace}[1]{{\rm Tr}\left\{ #1 \right\}}
 \newcommand{\traceB}[1]{{\rm Tr_B}\left\{ #1 \right\}}
 
 \newcommand{\abs}[1]{{\left| #1 \right|}}

 \newcommand{\sgn}[1]{{\rm sgn}\left( #1 \right)}

 \newcommand{\HS}{H_{\rm S}}
 \newcommand{\HB}{H_{\rm B}}
 \newcommand{\HI}{H_{\rm SB}}
 \newcommand{\RS}{\rho_{\rm S}}
 \newcommand{\RB}{\bar{\rho}_{\rm B}}
 \newcommand{\NS}{N_{\rm S}}
 \newcommand{\NB}{N_{\rm B}}
 \newcommand{\nb}{\rm n} 
 \newcommand{\fe}{{\rm f}} 

\begin{document}

\title{Quantum Equilibration under Constraints and Transport Balance}

\author{Gernot Schaller}\email{gernot.schaller@tu-berlin.de}

\affiliation{Institut f\"ur Theoretische Physik, Technische Universit\"at Berlin, Hardenbergstr. 36, 10623 Berlin, Germany}

\begin{abstract}
For open quantum systems coupled to a thermal bath at inverse temperature $\beta$,
it is well known that under the Born-, Markov-, and secular approximations the system
density matrix will approach the thermal Gibbs state with the bath inverse temperature $\beta$.
We generalize this to systems where there exists a conserved quantity (e.g., the total particle number), 
where for a bath characterized by inverse temperature $\beta$ and chemical potential $\mu$ we find 
equilibration of both temperature and chemical potential.
For couplings to multiple baths held at different temperatures and different chemical potentials, 
we identify a class of systems that equilibrates according to a single hypothetical average but in general 
non-thermal bath, which may be exploited to generate desired non-thermal states.
Under special circumstances the stationary state may be again be described by a unique Boltzmann factor.
These results are illustrated by several examples.
\end{abstract}

\pacs{05.60.Gg,  
03.65.Yz 
}

\maketitle

Thermalization is a classical phenomenon:
Coupling two materials at different temperature will lead to 
equilibration at some intermediate temperature -- depending on the
heat capacities of the constituents.
Especially when one piece is significantly larger than the other,
the temperature of the larger piece will hardly change, such that it may
be understood as a heat bath.
In contrast, the temperature of the smaller piece will simply approach the 
bath temperature in that limit.

The dynamics of open quantum systems that are coupled to a thermal bath is however
more difficile~\cite{srednicki1994a,linden2009a}.
A powerful tool to describe the evolution of such systems in various limits is the
quantum master equation~\cite{breuer2002,schlosshauer2008}:
A first order differential equation -- typically with constant coefficients -- describing 
the evolution of the system part of the density matrix.
As the derivation of an exact master equation is impossible in most cases, one has to
rely on perturbative schemes.
In such schemes, it is often already a challenge to preserve the fundamental
properties of the density matrix such as its trace, its self-adjointness, and its positive 
semidefiniteness.
Starting from microscopic models, especially the last property is often hard 
to fulfill, as for master equations with constant coefficients, preservation of 
positivity requires the dissipator to be of Lindblad~\cite{lindblad1976a} form.
Such Lindblad form dissipators are generically derived in the singular coupling limit~\cite{gorini1978a}, 
the weak-coupling limit -- also termed Born-Markov-secular~\cite{wichterich2007a} (BMS) approximation --
and in coarse-graining schemes~\cite{lidar2001a}.
Within the BMS approximation, thermalization of the system and equilibration of the systems 
temperature with that of the bath have been proven~\cite{kossakowski1977a}.
However, some baths are not only described by a temperature, but may also equilibrate under
further side constraints -- typically modeled by a chemical potential.
When we consider couplings to multiple baths held at different temperatures~\cite{wichterich2007a,segal2008a} and/or 
different chemical potentials~\cite{stoof1996a,gurvitz1996a}, we have the generic situation for 
transport~\cite{foerster2008a} from one reservoir through the system to another reservoir, 
which may be used to generate interesting 
non-equilibrium stationary states in the system.

This paper is organized as follows:
Having introduced the terminology in Sec.~\ref{Spreliminaries} we show how conserved quantities
lead to additional properties of the dampening coefficients in Sec.~\ref{Sconserved}.
The case of a single bath is discussed in Sec.~\ref{SSsingle}, followed by a discussion of multiple 
baths in Sec.~\ref{SSmultiple}.
We derive general statements on the resulting non-equilibrium stationary state for master equations
that are tridiagonal in the system energy eigenbasis in Sec.~\ref{SSladder}.
The conditions under which such a stationary state may still appear thermal are discussed in Sec.~\ref{SStrivial}.
Finally, the results are demonstrated with a number of examples in Sec.~\ref{Sexamples}.


\section{Preliminaries}\label{Spreliminaries}

We will consider a large closed quantum system with the total Hamiltonian
\bea
H = \HS+\HB+\HI\,,
\eea
where $\HS$ and $\HB$ act only on the system and bath parts, respectively, and $\HI$ mediates 
a coupling.
The latter may generally be decomposed as~\cite{breuer2002}
\bea\label{Eintdecomp}
\HI = \sum_{\alpha=1}^M A_\alpha \otimes B_\alpha
\eea
with $M$ hermitian system ($A_\alpha=A_\alpha^\dagger$) and bath ($B_\alpha=B_\alpha^\dagger$) coupling operators.
By convention, the system coupling operators may be chosen traceless and orthonormal
$\trace{A_\alpha A_\beta}=\delta_{\alpha\beta}$.
For example, for an $N$-dimensional system Hilbert space one may use  
the $M=N^2-1$ generators of the symmetry group SU($N$) for the system coupling operators.

Under the Born, Markov, and secular (BMS) approximations~\cite{breuer2002},
and assuming that the bath is kept in an equilibrium state $\RB$
with the properties $\traceB{B_\alpha \RB}=0$ as well as $\left[\HB,\RB\right]=0$,
one derives a master equation of Lindblad form for the system density matrix $\RS$.
In the system energy eigenbasis $\HS \ket{a}\equiv E_a \ket{a}$
it assumes the form~\cite{schaller2008a}
\bea\label{Elindblad_bms}
\dot{\RS} &=& -i \left[\HS+\sum_{ab} \tilde \sigma_{ab} \ket{a}\bra{b}, \RS(t)\right]\nn
&&+\sum_{abcd} \tilde \gamma_{ab, cd} \Big[\Big(\ket{a}\bra{b}\Big) \RS(t) \Big(\ket{c}\bra{d}\Big)^\dagger\nn
&&-\frac{1}{2}\left\{\Big(\ket{c}\bra{d}\Big)^\dagger \Big(\ket{a}\bra{b}\Big), \RS(t)\right\} \Big]\,,
\eea
where $\tilde\sigma_{ab}=\tilde\sigma_{ba}^*$ defines the unitary action of decoherence 
(also denoted Lamb-shift~\cite{breuer2002} or exchange field~\cite{braun2004a}) and the
dampening coefficients $\tilde\gamma_{ab, cd}$ describe the non-unitary (dissipative) 
terms due to the interaction with the reservoir.
The net effect of the secular approximation mentioned above is that these coefficients 
may vanish when some transition frequencies are not matched 
\bea\label{Ecoefficients}
\tilde \sigma_{ab} &=& \frac{1}{2i} \sum_c \sum_{\alpha\beta} \sigma_{\alpha\beta}(E_a-E_c) 
\delta_{E_b, E_a}\times\nn
&&\times \bra{c}A_{\alpha}\ket{a}^* \bra{c} A_{\beta} \ket{b}\,,\nn
\tilde \gamma_{ab, cd} &=& \sum_{\alpha\beta} \gamma_{\alpha\beta}(E_b-E_a) \delta_{E_d-E_c, E_b-E_a}\times\nn
&&\times \bra{a}A_{\beta}\ket{b} \bra{c} A_{\alpha} \ket{d}^*\,,
\eea
which is formally expressed by the Kronecker-$\delta$ symbols 
(but see \cite{schaller2008a,schaller2009a} for a Lindblad-form 
coarse-graining approach circumventing the secular approximation):
The Lamb-shift Hamiltonian for example will only act within the subspace of energetically degenerate states.
Note that when the spectrum of the system Hamiltonian is non-degenerate, Eq.~(\ref{Elindblad_bms}) may be 
simplified into a rate equation system (which is independent on the Lamb-shift) for the diagonals of 
$\RS$ in the energy eigenbasis.
In the dampening coefficients, the functions
\bea\label{Eft}
\gamma_{\alpha\beta}(\omega) &\equiv& 
\int\limits_{-\infty}^{+\infty} C_{\alpha\beta}(\tau) e^{+i \omega \tau} d\tau\,,\nn
\sigma_{\alpha\beta}(\omega) &\equiv&
\int\limits_{-\infty}^{+\infty} C_{\alpha\beta}(\tau) \sgn{\tau} e^{+i \omega \tau} d\tau
\eea
are even ($\gamma_{\alpha\beta}$) and odd ($\sigma_{\alpha\beta}$) Fourier transforms of the bath correlation functions
\bea
{\cal C}_{\alpha\beta}(\tau)\equiv
\traceB{e^{+i \tau \HB} B_\alpha e^{-i \tau \HB} B_\beta \RB}\,.
\eea
The bath correlation functions have many interesting analytic properties~\cite{breuer2002}.
For example, when the bath is held at a thermal equilibrium state (canonical ensemble)
\bea
\RB = \frac{e^{-\beta \HB}}{\traceB{e^{-\beta \HB}}}\,,
\eea
one can easily verify~\cite{breuer2002} the 
Kubo-Martin-Schwinger~\cite{kubo1957a,martin1959a,kubo1966a} (KMS) condition
${\cal C}_{\alpha\beta}(\tau) = {\cal C}_{\beta\alpha}(-\tau-i\beta)$.
Since the bath correlation functions are analytic in the lower complex half plane, 
the Fourier transform of the KMS condition reads
\bea\label{Ekms}
\gamma_{\alpha\beta}(-\omega) = e^{-\beta\omega} \gamma_{\beta\alpha}(+\omega)\,,
\eea
and can be used to prove~\cite{kossakowski1977a} that the equilibrated Gibbs state
\bea\label{Egibbs}
\bar\RS = \frac{e^{-\beta \HS}}{\trace{e^{-\beta \HS}}}
\eea
is a stationary state of Eq.~(\ref{Elindblad_bms}).


\section{Conserved Quantities}\label{Sconserved}

Now assume that there exists a conserved quantity $N=N_S+N_B$, where $N_S$ and $N_B$ act only on system and bath, 
respectively.
That is, we assume that $\left[\HS,N_S\right]=0$, $\left[\HB,N_B\right]=0$, and
$\left[\HI, N_S+N_B\right]=0$, such that nontrivial evolution arises via 
\mbox{$\left[\HI, N_S\right] \neq 0 \neq \left[\HI, N_B\right]$}.
The conservation laws imply the identity 
\mbox{$\HI=e^{-\kappa (\NS+\NB)} \HI e^{+\kappa(\NS+\NB)}$}.
Acting on this expression with $e^{+\kappa \NS}[\ldots]e^{-\kappa \NS}$ yields with
Eq.~(\ref{Eintdecomp}) the identity
\bea\label{Econservationlaw}
\sum_\alpha \left(e^{+\kappa\NS} A_\alpha e^{-\kappa\NS}\right)\otimes B_\alpha 
=\nn
\sum_\alpha A_\alpha \otimes \left(e^{-\kappa\NB}B_\alpha e^{+\kappa\NB}\right)\,,
\eea
such that effectively a transformation of the form 
$e^{+\kappa \NS}[\ldots]e^{-\kappa \NS}$ on a system coupling operator is mapped 
into a transformation of the form
$e^{-\kappa \NB}[\ldots]e^{-\kappa \NB}$ on the corresponding bath coupling operator.
We will show in the following that this identity leads to additional properties of
the dampening coefficients in Eq.~(\ref{Elindblad_bms}), when the bath density
matrix is assumed to be in the grand-canonical equilibrium state with chemical potential $\mu$
\bea\label{Egibbsgcan}
\RB = \frac{e^{-\beta \left(\HB-\mu \NB\right)}}{\traceB{e^{-\beta \left(\HB-\mu \NB\right)}}}\,.
\eea
Note that -- depending on the spectrum of $\HB$ -- normalizability of $\RB$ may impose 
constraints on the chemical potential, compare 
Sec.~\ref{SSoscillators}, Sec.~\ref{SSspinboson}, and Sec.~\ref{SSmixed}.

Evidently, as $\left[\HS,N_S\right]=0$, we may and will in the following choose 
$\ket{a}$ to be the common eigenbasis of the two operators with 
$\HS \ket{a} \equiv E_a \ket{a}$ and $N_S \ket{a} \equiv N_a \ket{a}$.

When we multiply the Lamb-shift coefficients $\tilde \sigma_{ij}$ by a factor of the form
$e^{+\beta\mu (N_i-N_j)}$, we may use Eq.~(\ref{Ecoefficients}) to replace the eigenvalues by operators,
such that the system operators in Eq.~(\ref{Ecoefficients}) are rotated.
Then, the identity~(\ref{Econservationlaw}) with Eq.~(\ref{Eft}) can be used to transfer the pseudo-rotation
to the bath correlation functions.
Finally, we may use the invariance of the trace over the bath degrees of freedom under cyclic permutations
and $\left[\NB,\RB\right]=0$ to see that 
\bea\label{Erelation1}
\tilde\sigma_{ij}e^{+\beta\mu (N_i-N_j)}  = \tilde\sigma_{ij}\,,
\eea
i.e., the Lamb shift Hamiltonian only acts on states with both degenerate energy and particle number.
An analog calculation for the dissipative coefficients $\tilde\gamma_{ab,cd}$ reveals the identity
\bea\label{Erelation2}
\tilde\gamma_{aj,ai} e^{+\beta\mu(N_i-N_j)} = \tilde\gamma_{aj,ai}.
\eea
When we consider additional thermal Boltzmann factors, one needs to change the integration path in the
Fourier transform~(\ref{Eft}) -- using that the bath correlation functions are analytic -- to show that 
the balance relation
\bea\label{Erelation3}
\tilde\gamma_{ia,ja} e^{-\beta (E_a-E_i)} e^{+\beta\mu(N_a-N_j)} = \tilde \gamma_{aj,ai}
\eea
holds.
Such relations are termed fluctuation theorems~\cite{esposito2009a}.
Specifically, relations (\ref{Erelation1}), (\ref{Erelation2}), and (\ref{Erelation3}) generalize 
the KMS condition ($\mu=0$) for the quantum master equation in Eq.~(\ref{Ekms})
to systems with a conserved quantity.


\section{Stationary State}\label{Sstatstate}


\subsection{Single Reservoir}\label{SSsingle}

The matrix elements of the grand-canonical Gibbs state~(\ref{Egibbsgcan}) read
\bea\label{Egibbs2}
\bar{\rho}_{ij} = \frac{\bra{i}e^{-\beta \left(\HS-\mu\NS\right)}\ket{j}}{Z}
=\frac{\delta_{ij} e^{-\beta(E_i-\mu N_i)}}{Z}\,,
\eea
where $Z=\trace{e^{-\beta \left(\HS-\mu\NS\right)}}$ denotes the normalization.
For such a diagonal density matrix, the time-evolution of off-diagonal matrix elements
may in principle still be influenced by the diagonals, as Eq.~(\ref{Elindblad_bms}) reduces to
\bea\label{Ebmsstat}
\dot{\bar{\rho}}_{ij} &=& -i \tilde\sigma_{ij} \left(\bar\rho_{jj}-\bar\rho_{ii}\right)
+\sum_a \tilde\gamma_{ia,ja} \bar\rho_{aa}\nn
&&- \frac{1}{2} \sum_a \tilde\gamma_{aj,ai} \left(\bar\rho_{ii}+\bar\rho_{jj}\right)\,.
\eea
To show stationarity of the Gibbs state~(\ref{Egibbs2}),
it is convenient to distinguish different cases:
\begin{enumerate}
\item[a.] 
Trivially, when $i \neq j$ and also $E_i \neq E_j$ in Eq.~(\ref{Ebmsstat}), we have 
$\dot{\bar{\rho}}_{ij}=0$, since all coefficients simply vanish, cf. Eq.~(\ref{Ecoefficients}).
\item[b.]
When $i = j$ and evidently also $E_i = E_j$,  
Eq.~(\ref{Ebmsstat}) reduces to the rate equation system
\bea
\dot{\bar{\rho}}_{ii} &=&
+\sum_a \tilde\gamma_{ia,ia} \bar\rho_{aa} - \sum_a \tilde\gamma_{ai,ai} \bar\rho_{ii}=0\,,
\eea
which vanishes due to the detailed balance relation~(\ref{Erelation3}), evaluated for $i=j$.
\item[c.]
For degenerate energy levels we have the additional possibility that $i \neq j$ but $E_i=E_j$.
Cancellation of the Lamb-shift terms in~(\ref{Ebmsstat}) results from Eq.~(\ref{Erelation1}).
Showing that the remaining dissipative terms also vanish amounts to
\bea
0 &=& \sum_a \tilde\gamma_{ia,ja} e^{-\beta(E_a-E_i)}e^{+\beta\mu(N_a-N_j)}\nn
&&-\frac{1}{2} \sum_a \gamma_{aj,ai} \left[e^{+\beta\mu(N_i-N_j)}+1\right]\,,
\eea
which is directly evident from relations~(\ref{Erelation2}) and~(\ref{Erelation3}).
\end{enumerate}
To summarize, we have shown that the state~(\ref{Egibbs2}) is a stationary state of the 
quantum master equation~(\ref{Elindblad_bms}), when the reservoir density matrix is of the form~(\ref{Egibbsgcan}).
Generally of course, the existence of further stationary states is possible,
but for an ergodic~\cite{breuer2002} evolution the BMS approximation scheme for a single
reservoir leads to 
{\em equilibration of both temperature and chemical potential}.
This equilibration has been noted earlier for specific examples~\cite{schaller2009b,znidaric2010a,rigol2008a}
and has been used quite generally for systems of rate equations~\cite{esposito2007a}.
Here however, we have a rigorous proof for the quantum master equation.


\subsection{Multiple Reservoirs}\label{SSmultiple}

When the system of interest is is not only coupled to a single, but multiple ($K$) reservoirs
\bea\label{Ehintdecomp}
\HI = \sum_\alpha A_\alpha \otimes \sum_{k=1}^K B_\alpha^{(k)}\,,\;\; \HB = \sum_{k=1}^K \HB^{(k)}\,,
\eea
where varying coupling strengths are absorbed in the $B_\alpha^{(k)}$ operators and the independent reservoirs
are characterized by different inverse temperatures $\beta^{(k)}$ and different chemical potentials $\mu^{(k)}$
\bea\label{Ebathproduct}
\RB = \bigotimes\limits_{k=1}^K 
\frac{e^{-\beta^{(k)} \left(\HB^{(k)}-\mu^{(k)} N_B^{(k)}\right)}}
{{\rm Tr_B^{(k)}}\left\{e^{-\beta^{(k)} \left(\HB^{(k)}-\mu^{(k)} N_B^{(k)}\right)}
\right\}}\,,
\eea 
much less is known about the resulting stationary state~\cite{saito2008a}.
A decomposition of the interaction Hamiltonian in the form of Eq.~(\ref{Ehintdecomp}) with identical system coupling
operators for each bath is always possible, as we have chosen the $A_\alpha$ operators to form a complete 
basis set for hermitian operators in the system Hilbert space.
We assume that some interaction Hamiltonians may obey a conserved quantity $N^{(k)}\equiv N_S+N_B^{(k)}$, where
$\left[N_B^{(k)},\HB^{(k)}\right]=0=\left[N_S+N_B^{(k)},\sum_\alpha A_\alpha \otimes B_\alpha^{(k)}\right]$.
Evidently, the form of Eq.~(\ref{Elindblad_bms}) remains invariant with
\bea\label{Ecombined}
\tilde\gamma_{ab,cd} = \sum_{k=1}^K \tilde\gamma_{ab,cd}^{(k)}\,,\qquad
\tilde\sigma_{ab} = \sum_{k=1}^K \tilde\sigma_{ab}^{(k)}\,,
\eea
where $\tilde\gamma_{ab,cd}^{(k)}$ and $\tilde\sigma_{ab}^{(k)}$ describe the dissipation and Lamb-shift, 
respectively, due to the $k$-th bath only.
Accordingly, each bath yields separate detailed balance conditions of the form of 
Eqns.~(\ref{Erelation1}),~(\ref{Erelation2}), and~(\ref{Erelation3}).
In general, this will lead to a non-equilibrium stationary state.


\subsection{Rate Equations for Ladder Spectra}\label{SSladder}

Quite general statements on the resulting non-equilibrium stationary state may be 
obtained for master equations that assume tri-diagonal form in an $N$-dimensional energy and number eigenbasis
(with $\HS \ket{m}=E_m \ket{m}$ and $N_S\ket{m}=N_m\ket{m}$), 
where $\rho_m\equiv \bra{m}\RS\ket{m}$ and we assume ordering with respect
to the (quasi-)particle numbers in the system ($N_{m+1}-N_m=1$)
\bea\label{Erateex1}
\dot \rho_m &=& \tilde\gamma_{m,m-1} \rho_{m-1} + \tilde\gamma_{m,m+1} \rho_{m+1}\nn
&&- \left[\tilde\gamma_{m-1,m}+\tilde\gamma_{m+1,m}\right]\rho_m\,.
\eea
In this basis, the populations evolve independently from the coherences (which we assume to decay), 
and transitions between populations are mediated by single particle tunneling processes.
Note however, that even an effective rate equation system of the form~(\ref{Erateex1}) may keep
genuine quantum properties as the eigenstates $\ket{m}$ themselves may e.g.\ be entangled between
different sub-parts of the system.
At the boundaries $m_1$ and $m_2$ of the spectrum (where $N_{m_1}=0$ or $N_{m_2}=N$) the unphysical tunneling rates vanish, for
example we have $\dot\rho_{m_1}=+\tilde\gamma_{m_1,m_1+1}\rho_{m_1+1}-\tilde\gamma_{m_1+1,m_1} \rho_{m_1}$.
Computing the stationary state and using the results from the boundary, this implies that for all $m$, it has to satisfy
\bea\label{Econdstat}
\frac{\bar\rho_{m+1}}{\bar\rho_m} = \frac{\tilde\gamma_{m+1,m}}{\tilde\gamma_{m,m+1}}\,.
\eea
Together with the trace condition $\sum_m \bar\rho_m=1$ this completely defines the stationary state.
In addition, we assume that (compare Sec.~\ref{Sexamples} for examples) the dampening coefficients associated 
with the $k^{\rm th}$ bath have the decomposition
\bea\label{Edecomposition}
\tilde\gamma_{m,m+1}^{(k)} &=& g_m G^{(k)}(\omega_{m+1,m}) \left[1 \pm F_\pm^{(k)}(\omega_{m+1,m})\right]\,,\nn
\tilde\gamma_{m+1,m}^{(k)} &=& g_m G^{(k)}(\omega_{m+1,m}) F_\pm^{(k)}(\omega_{m+1,m})\,,
\eea
where $\omega_{m+1,m}\equiv E_{m+1}-E_m$, and $g_m$, $G^{(k)}(\omega)$ and 
$F_\pm^{(k)}(\omega)$ contain the details of the system, the coupling, and the thermal properties
of the bath, respectively.
Using this factorization in the local balance condition (\ref{Erelation3}) leads with $N_{m+1}-N_m=1$ to
$\tilde\gamma_{m,m+1}^{(k)} e^{-\beta^{(k)} \omega_{m+1,m}} e^{+\beta^{(k)} \mu^{(k)}}=\tilde\gamma_{m+1,m}^{(k)}$,
which is automatically fulfilled for fermionic
\bea\label{Efermi}
F_-^{(k)}(\omega) &=& \frac{1}{e^{\beta^{(k)}\left(\omega-\mu^{(k)}\right)}+1}
\equiv \fe^{(k)}(\omega)
\eea
or bosonic
\bea\label{Ebose}
F_+^{(k)}(\omega) &=& \frac{1}{e^{\beta^{(k)}\left(\omega-\mu^{(k)}\right)}-1}
\equiv \nb^{(k)}(\omega)
\eea
baths.
It is obvious from Eq.~(\ref{Econdstat}) that for coupling to a single bath, neither $g_m$ nor 
the energy dependence of the tunneling rate $G^{(k)}(\omega)$ do affect the stationary state -- 
only the transient relaxation dynamics will be changed.
However, when the system is coupled to multiple baths obeying Eqns.~(\ref{Edecomposition}), it is easy to
show that the stationary state -- as characterized by Eq.~(\ref{Econdstat})
\bea\label{Estatmatrate}
\frac{\bar\rho_{m+1}}{\bar\rho_m} &=& 
\frac{\sum_k G^{(k)}(\omega_{m+1,m}) F_\pm^{(k)}(\omega_{m+1,m})}
{\sum_k G^{(k)}(\omega_{m+1,m})\left[1\pm F_\pm^{(k)}(\omega_{m+1,m})\right]}\nn
&=& \frac{\bar{F}_\pm(\omega_{m+1,m})}{1\pm \bar{F}_\pm(\omega_{m+1,m})}
\eea
is the same as one for a hypothetical single structured bath with the weighted average occupation function
\bea\label{Edistaverage}
\bar{F}_\pm(\omega) \equiv \frac{\sum_k G^{(k)}(\omega) F_\pm^{(k)}(\omega)}{\sum_k G^{(k)}(\omega)}.
\eea
For a single ($K=1$) bath of either fermionic or bosonic nature the ratio in~(\ref{Estatmatrate}) reduces
to the conventional Boltzmann factor $\bar\rho_{m+1}/\bar\rho_m=e^{-\beta^{(1)}(\omega_{m+1,m}-\mu^{(1)})}$,
such that we will term it generalized Boltzmann factor further-on.
Evidently, the energy dependence of the tunneling rates enters the average occupation 
function~(\ref{Edistaverage}) and may therefore be 
used to tune the resulting non-equilibrium stationary state.
For example, whereas the canonical Gibbs state~(\ref{Egibbs}) will for finite temperature always favor
the ground state and the grand-canonical Gibbs state~(\ref{Egibbsgcan}) may favor states with a certain 
particle number, it is here possible e.g.\ to select multiple states of interest.
Note however, that with using only bosonic baths (with $\nb(\omega)\ge0$) it is not possible to 
achieve e.g. $\bar\rho_{m+1}>\bar\rho_m$, which is in stark contrast to fermionic baths, where this only 
requires $\bar{F}_-(\omega_{m+1,m})>1/2$.

To illustrate this idea let us for simplicity parameterize the tunneling rates phenomenologically 
(but see e.g.~\cite{elattari2000a,zedler2009a} for a microscopic justification)
by a Lorentzian shape
\bea\label{Elorentz}
\Gamma^{(k)}(\omega) = \frac{\Gamma_k \delta_k^2}{(\omega-\bar\omega_k)^2+\delta_k^2}
\eea
with maximum rate $\Gamma_k$ at frequency $\bar\omega_k$ and width $\delta_k$.
\begin{figure}
\includegraphics[height=6cm,clip=true]{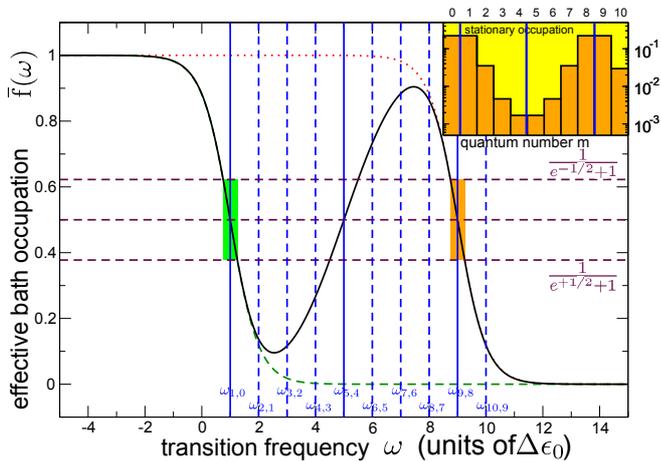}
\caption{\label{Ffermi_structure}
(Color Online)
Effective bath occupation (solid black) in Eq.~(\ref{Edistaverage}) generated by two fermionic baths (dotted red and dashed green
curves, chemical potentials $\mu^{(k)}$ and temperatures $\left[\beta^{(k)}\right]^{-1}$ are given by the 
center and width of the shaded boxes, respectively) with Lorentzian tunneling rates of form~(\ref{Elorentz}).
The detuning of the tunneling rates is adjusted such that the average hypothetical occupation function $\bar\fe(\omega)$
is not monotonically decaying.
Then, the threshold $1/2$ is passed several times (vertical solid blue lines) within the transition frequency
window probed by the system.
This will lead to an increasing population $\rho_{m+1}>\rho_m$ when $\bar\fe(\omega_{m+1,m})>1/2$ and to a decreasing
population $\rho_{m+1}<\rho_m$ when $\bar\fe(\omega_{m+1,m})<1/2$ (compare the inset).
Such multi-modal distributions are clearly non-thermal, i.e., they cannot be generated by a single thermal bath by
adjusting temperature and chemical potential.
Parameters have been chosen as $\Delta\epsilon_0\beta^{(1)}=\Delta\epsilon_0\beta^{(2)}=2$, 
$\mu^{(1)}=\Delta\epsilon_0$, $\mu^{(2)}=9\Delta\epsilon_0$, $\Gamma_1=\Gamma_2$, 
$\delta_1=\delta_2=0.1\Delta\epsilon_0$, and $\bar\omega_1=\Delta\epsilon_0$ with $\bar\omega_2=9\Delta\epsilon_0$.
The shown equidistant transition frequencies (vertical solid and dashed blue lines) correspond to eigenvalues that scale 
quadratically with $m$
and may be realized by the example in Sec.~\ref{SSelectronic} 
by using $N=10$, $\epsilon=\Delta\epsilon_0$, $U=\Delta\epsilon_0$, and $T=0$.
}
\end{figure}
When the average occupation function at the transition frequency $\omega_{m+1,m}$ between states $\ket{m}$ and
$\ket{m+1}$ exceeds 1/2, the population of the states will increase $\rho_{m+1}>\rho_m$, whereas the opposite is true 
for $\bar{\fe}(\omega_{m+1,m})<1/2$, see Fig.~\ref{Ffermi_structure}.
Depending on the number and the thermal properties used baths, the resulting hypothetical average 
distribution (\ref{Edistaverage}) may assume quite arbitrary shapes, such that more sophisticated 
statistical stationary mixtures than in Fig.~\ref{Ffermi_structure} are conceivable.


\subsection{Trivial dynamic equilibria}\label{SStrivial}

It has been observed for special systems~\cite{segal2008a} that the equilibrium state had a thermal form
with some average temperature even though the system of interest was coupled to more than just
a single thermal bath with different temperatures.
For the ladder-like systems discussed in the previous section, it is obvious that to describe
the stationary state with just two parameters $\bar\beta$ and $\bar\mu$ requires Eqns.~(\ref{Estatmatrate})
to be compatible with
\bea\label{Esingleboltz}
\frac{\bar\rho_{m+1}}{\bar\rho_m} 
&=& \frac{\bar{F}_\pm(\omega_{m+1,m})}{1\pm \bar{F}_\pm(\omega_{m+1,m})}
\stackrel{!}{=} e^{-\bar\beta(\omega_{m+1,m}-\bar\mu)}\,,
\eea
where the effective average occupation is defined in Eq.~(\ref{Edistaverage}).
This can be achieved in several possible ways:

Firstly, for a three-dimensional system Hilbert space ($N=2$ in the previous section), 
we will only have two transition frequencies and
accordingly only two equations of the above form.
These two equations would uniquely fix the parameters $\bar\beta$ and $\bar\mu$.
Note however, that then in general $\bar\beta<0$ is possible, which corresponds to population inversion 
(i.e., favoring for $\bar\mu=0$ the most excited state)
Evidently, this forbids the interpretation of the parameter $\bar\beta$ as
inverse temperature.

Alternatively, for $N>2$, Eqns.~(\ref{Estatmatrate}) may still be linearly dependent, as is the case when
the system Hamiltonian provides only a single transition frequency (e.g., a harmonic oscillator).
In an approximate sense, linear dependence may also be generated when the average occupation function
$\bar{F}_\pm(\omega)$ is essentially flat in the transition frequency window probed by the system.
Assuming only a single transition frequency $\Omega$, such that the tunneling rates are described by a single number $G^{(k)}(\Omega)\equiv\Gamma_k$,
and vanishing chemical potentials $\mu^{(k)}=0=\bar\mu$ in the low-energy limit $\beta^{(k)} \Omega \ll 1$, Eqn.~(\ref{Estatmatrate}) defines average temperatures
for bosons and fermions
\bea\label{Etempaverage}
\bar{T}_{\rm Bose} &\approx& \sum_k \frac{\Gamma_k}{\Gamma} T^{(k)}\,,\nn
\frac{1}{\bar{T}_{\rm Fermi}} &\approx& \sum_k \frac{\Gamma_k}{\Gamma} \frac{1}{T^{(k)}}
\eea
with $\Gamma\equiv\sum_\ell \Gamma_\ell$, 
that correspond to the weighted arithmetic (bosons) and weighted harmonic (fermions) mean.
The bosonic average temperature does well resemble the Richmann mixing formula and has been found previously~\cite{segal2008a}.
However, it should be noted that this is different from the so-called classical (high-energy) limit $\beta^{(k)} \Omega \gg 1$, for which one obtains an arithmetic
mean of the Boltzmann factors from Eq.~(\ref{Estatmatrate})
\bea
e^{-\bar\beta\Omega} \approx \sum_k \frac{\Gamma_k}{\Gamma} e^{-\beta^{(k)} \Omega}
\eea
for -- as expected -- both fermionic and bosonic baths.


\section{Examples}\label{Sexamples}

For a single reservoir the equilibration of both temperature and chemical 
potential has been observed for interacting double dots~\cite{schaller2009b} in the
BMS approximation.
Therefore, we only give examples to illustrate the results in 
Sec.~\ref{SSmultiple}, Sec.~\ref{SSladder}, and Sec.~\ref{SStrivial}.
Naturally, in case of only a single coupling bath, the case of Sec.~\ref{SSsingle} 
is also reproduced here.

\subsection{Homogeneous Electronic Nanostructure}\label{SSelectronic}

Consider a nanostructure with $N$ homogenous electronic sites (we neglect the spin)
\bea
\HS=\epsilon \sum_{i=1}^N d_i^\dagger d_i+\frac{U}{2} \sum_{i\neq j} d_i^\dagger d_i d_j^\dagger d_j 
+ T \sum_{i \neq j} d_i^\dagger d_j
\eea
with single-particle energy $\epsilon$, Coulomb interaction $U$, and
hopping term $T$ that are all assumed completely isotropic.
The permutational symmetry suggests to reduce the problem to the symmetrized subspace
with the basis
\mbox{$\ket{N,m+1}\equiv \frac{1}{\sqrt{(N-m)(m+1)}} D^\dagger \ket{N,m}$} with 
\mbox{$0 \le m \le N$} denoting the number of electrons in the system, 
where $D\equiv\sum_{i=1}^N d_i$ and $\ket{N,0}=\ket{0,\ldots,0}$ represents the $N$ particle Fock space vacuum.
Clearly, both $\HS$ and $N_S\equiv\sum_i d_i^\dagger d_i$ are diagonal in this basis.
For $N=1$ we recover the single resonant level~\cite{haug2008}, but for 
for $N>1$, the spectrum of the system Hamiltonian becomes nontrivial.
The eigenvalues of the symmetric subspace read $E_m=m\epsilon+m(m-1) U/2 + m(N-m) T$, such that the
spectrum may become equidistant when $U=2T$ with 
$\omega_{m+1,m}\to\epsilon+(N-1) T$ -- which still admits a strongly interacting model.

At first we assume that the nanostructure is only coupled to a single lead
$\HB=\sum_k \epsilon_k c_k^\dagger c_k$ at temperature $\beta$ and chemical potential $\mu$ via the tunneling 
Hamiltonian~\footnote{For fermionic system and bath operators that -- strictly speaking -- must anti-commute, 
such a tensor product decomposition may be obtained by using the Jordan-Wigner transform. 
Fermions separately defined on the system and bath Hilbert space may then be re-introduced using
an inverse Jordan-Wigner transform on the respective Hilbert space only~\cite{schaller2009b}.
}
$\HI = D \otimes \sum_k g_k c_k^\dagger + D^\dagger \otimes \sum_k g_k^* c_k$, 
where $g_k$ represents a frequency-dependent coupling constant, and
$c_k$ are fermionic annihilation operators acting on the lead Hilbert space.
The conserved quantity is composed from $N_S$ and
$N_B = \sum_k c_k^\dagger c_k$.
We may write the interaction Hamiltonian also as
\mbox{$\HI = A_1 \otimes B_1 + A_2 \otimes B_2$},
with the hermitian and trace-orthogonal (not-normalized) system coupling operators
$A_1 = \left( D^\dagger+D\right)$ and
$A_2 = i\left(D^\dagger-D\right)$, 
and the associated hermitian bath coupling operators
\mbox{$B_1 = \sum_k\left( g_k c_k^\dagger + g_k^* c_k\right)/2$} and 
\mbox{$B_2 = i \sum_k\left( g_k c_k^\dagger - g_k^* c_k\right)/2$}.
For a bath in thermal equilibrium with inverse temperature $\beta$ and chemical potential $\mu$, such that
its density matrix is given by Eq.~(\ref{Egibbsgcan}), we obtain for Fourier transforms~(\ref{Eft}) of the bath 
correlation functions
$\gamma_{11}(\omega)=\gamma_{22}(\omega)=\Gamma(+\omega)\left[1-\fe(+\omega)\right]/4+\Gamma(-\omega)\fe(-\omega)/4$
and
$\gamma_{12}(\omega)=\gamma_{21}^*(\omega)=i \Gamma(+\omega)\left[1-\fe(+\omega)\right]/4
-i \Gamma(-\omega)\fe(-\omega)/4$,
where $\Gamma(\omega) \equiv 2\pi \sum_k \abs{g_k}^2 \delta(\omega-\epsilon_k)$ is the tunneling rate 
and the Fermi function $\fe(\omega)$ encodes the bath properties $\beta$ and
$\mu$, compare Eq.~(\ref{Efermi}).
Obviously, the Fourier-transform matrix of bath correlation functions has non-negative eigenvalues -- 
a consequence of Bochners theorem~\cite{breuer2002,reed1975}.
These lead to the non-vanishing dampening coefficients
\bea\label{Econstex1}
\tilde\gamma_{m,m+1} &=& (N-m)(m+1)\Gamma(\omega_{m+1,m})\left[1-\fe(\omega_{m+1,m})\right]\,,\nn
\tilde\gamma_{m+1,m} &=& (N-m)(m+1)\Gamma(\omega_{m+1,m})\fe(\omega_{m+1,m})\,,
\eea
where $\omega_{m+1,m}\equiv E_{m+1}-E_m$ and the factoring condition (\ref{Edecomposition}) is obviously fulfilled.
We obtain a rate equation of the form (\ref{Erateex1}) with $\rho_m \equiv \bra{N,m}\RS\ket{N,m}$, where 
the Lamb-shift terms are irrelevant, since we have by exploiting the permutational symmetry mapped our system
to a nondegenerate one.

Now we consider tunnel couplings to multiple baths with factorizing density matrices as in Eq.~(\ref{Ebathproduct}).
The form of Eq.~(\ref{Erateex1}) remains invariant, and we simply have 
$\tilde\gamma_{m,m\pm 1}=\sum_k \tilde\gamma_{m,m\pm 1}^{(k)}$, with different temperatures $\beta^{(k)}$ 
and chemical potentials $\mu^{(k)}$ entering the rates as in Eq.~(\ref{Econstex1}).
The general non-equilibrium steady state of Eq.~(\ref{Erateex1}) fulfills
\bea\label{Efactornboltz}
\frac{\bar \rho_{m+1}}{\bar \rho_m} = \frac{\sum_k \Gamma^{(k)}(\omega_{m+1,m}) \fe^{(k)}(\omega_{m+1,m})}
{\sum_k \Gamma^{(k)}(\omega_{m+1,m})\left[1-\fe^{(k)}(\omega_{m+1,m})\right]}
\eea
and is thus identical with the non-equilibrium steady state for coupling to a single hypothetical non-equilibrium 
bath with average non-thermal distribution of the form (\ref{Edistaverage}), compare also Fig.~\ref{Ffermi_structure}.


\subsection{Coupled oscillators}\label{SSoscillators}

We consider a harmonic oscillator $\HS=\Omega b^\dagger b$ coupled to
many others $\HB = \sum_k \omega_k b_k^\dagger b_k$ with positive eigenfrequencies
$\omega_k >0$ via quasi-particle tunneling
\mbox{$\HI=b \otimes \sum_k h_k b_k^\dagger + b^\dagger \otimes \sum_k h_k^* b_k$}.
The conserved quantity is composed from $N_S=b^\dagger b$ and $N_B=\sum_k b_k^\dagger b_k$.
Rewriting the interaction Hamiltonian in terms of hermitian operators, we obtain
$A_1 = (b^\dagger+b)$, $A_2=i(b^\dagger-b)$, 
$B_1 = \sum_k (h_k b_k^\dagger+h_k^* b_k)/2$, and $B_2=i\sum_k(h_k b_k^\dagger-h_k^* b_k)/2$.
The matrix elements of the Fourier transforms~(\ref{Eft}) of the bath correlation functions equate to
$\gamma_{11}(\omega)=\gamma_{22}(\omega)= 
1/4\left[+\Theta(+\omega)[1+\nb(+\omega)]\Gamma(+\omega)+\Theta(-\omega)\nb(-\omega)\Gamma(-\omega)\right]$
for the diagonals and 
$\gamma_{21}^*(\omega)=\gamma_{12}(\omega)=
i/4\left[+\Theta(+\omega)[1+\nb(+\omega)]\Gamma(+\omega)-\Theta(-\omega)\nb(-\omega)\Gamma(-\omega)\right]$
for the off-diagonals,
where $\Theta(\omega)$ denotes the Heaviside step function, 
$\Gamma(\omega)\equiv2\pi \sum_k \abs{h_k}^2 \delta(\omega-\omega_k)$ the quasi-particle tunneling rate, and 
$\nb(\omega)$ denotes the bosonic occupation number as defined in Eq.~(\ref{Ebose}).
The condition that $\mu < \min_k (\omega_k)$ grants positivity of all bath occupations.
In addition, it implies that $\left.\Gamma(\Omega)\right|_{\Omega<\mu}=0$, such that we assume also $\Omega>\mu$
throughout.
Accordingly, the Fourier transform matrix of the bath correlation functions is positive semidefinite at
all frequencies.
In the Fock space basis (where $b^\dagger \ket{n}=\sqrt{n+1}\ket{n+1}$), we obtain a rate
equation of the form~(\ref{Erateex1}), where $0\le n < \infty$.
However, since even the original eigenstates are non-degenerate, the dampening coefficients equate (with $\Omega>0$) to 
\bea
\tilde\gamma_{n,n+1} &=& (n+1) \Gamma(\Omega) \left[1+\nb(\Omega)\right]\,,\nn
\tilde\gamma_{n+1,n} &=& (n+1) \Gamma(\Omega) \nb(\Omega)\,,
\eea
which is also compatible with assumption~(\ref{Edecomposition}).
The resulting system is infinitely large, one may however, introduce a cutoff size $N_{\rm cut}$ and solve 
for the stationary state of the rate equations for finite $N_{\rm cut}$.
The ratio $\rho_{n+1}/\rho_n$ of two successive populations yields the desired Boltzmann factor,
which even happens to be independent of $N_{\rm cut}$.

For multiple baths, we obtain equilibration in a thermal state with the
unique generalized Boltzmann factor
\bea\label{Eboltzmann2}
\frac{\rho_{n+1}}{\rho_n} = 
\frac{\sum_k \Gamma^{(k)}(\Omega) \nb^{(k)}(\Omega)}{\sum_k \Gamma^{(k)}(\Omega)\left[1+\nb^{(k)}(\Omega)\right]}\,,
\eea
consistent with Eq.~(\ref{Estatmatrate}).
This generalized Boltzmann factor is the same that one would obtain for contact with a single hypothetical 
non-thermal bath at an average occupation compatible with Eq.~(\ref{Edistaverage}),
which in the high-temperature and $\mu^{(k)}\to0$
limit reduces to the bosonic average temperature in Eq.~(\ref{Etempaverage}).
This coincides well with the well-known temperature mixing formula and previous results~\cite{segal2008a}.


\subsection{Spin-Boson Model}\label{SSspinboson}

A variant of the spin-boson model coupled to a single bath has already been provided in 
Ref.~\cite{vogl2010a}, such that we here only generalize to $K \ge 2$ baths and non-vanishing chemical potentials.
We consider a large spin system $\HS=\Omega/2 J^z$ with $\Omega>0$, coupled to a bath of harmonic oscillators 
$\HB=\sum_k \omega_k b_k^\dagger b_k$ with $\omega_k>0$ via
$\HI=J^+ \otimes \sum_k h_k b_k^\dagger + J^- \otimes \sum_k h_k b_k$, where 
$J^\alpha\equiv\sum_{i=1}^N \sigma^\alpha_i$ and $\sigma^\pm=(\sigma^x \pm i \sigma^y)/2$.
The conserved quantity is given by 
$N_S+N_B=-J^z/2+\sum_k b_k^\dagger b_k$.
We impose the same conditions on the chemical potential(s) as before: $\mu < \min_k (\omega_k)$ and $\mu < \Omega$.
Choosing the system coupling operators as $A_1=J^x$ and $A_2=J^y$, 
the Fourier transforms~(\ref{Eft}) of the bath correlation function are identical to 
the previous section.
In order to calculate the dampening coefficients, permutational symmetry suggests to use the angular 
momentum basis $\ket{N/2,m}$ with $-N/2 \le m \le +N/2$.
Using that $J^\pm \ket{j,m}=\sqrt{j(j+1)-m(m\pm 1)} \ket{j,m\pm 1}$, we obtain a rate equation 
of the form~(\ref{Erateex1}) with the coefficients
\bea
\tilde\gamma_{m,m+1}&=&\left[\frac{N}{2}\left(\frac{N}{2}+1\right)-m(m+1)\right] 
\Gamma(\Omega) \left[1+\nb(\Omega)\right]\,,\nn
\tilde\gamma_{m+1,m}&=&\left[\frac{N}{2}\left(\frac{N}{2}+1\right)-m(m+1)\right] 
\Gamma(\Omega) \nb(\Omega)\,.
\eea
Solving this rate equation with multiple baths for its steady state yields a thermal
bath with the same unique generalized Boltzmann factor as Eq.~(\ref{Eboltzmann2}),
as predicted by Eq.~(\ref{Estatmatrate}).


\subsection{Mixed Spin Model}\label{SSmixed}

We consider a spin-1/2 system $\HS=\Omega/2 \sigma^z$ that is firstly coupled to a
bosonic bath $\HB^{(1)}=\sum_k \omega_k b_k^\dagger b_k$ via the dissipative coupling
$\HI^{(1)}=\sigma^+ \otimes \sum_k h_k b_k^\dagger + \sigma^- \otimes \sum_k h_k^* b_k$,
and secondly to a fermionic bath $\HB^{(2)}=\sum_k \epsilon_k c_k^\dagger c_k$ via the
coupling 
$\HI^{(2)}=\sigma^+ \otimes \sum_k g_k c_k^\dagger + \sigma^- \otimes \sum_k g_k^* c_k$.
Note that in contrast to the previous examples we do now consider two different baths from
the beginning.
The interaction Hamiltonians explicitly obeys the conserved quantity constructed from
$N_S=-\sigma^z/2$, $N_B^{(1)}=\sum_k b_k^\dagger b_k$, and $N_B^{(2)}=\sum_k c_k^\dagger c_k$.
Note that $\HI^{(2)}$ does not conserve the number of fermions, but such a model may 
represent scattering processes with a further fermionic bath that omitted from the description.
As before, we require that $\mu^{(1)} < \Omega$ and $\mu^{(1)}<\min_k(\omega_k)$.
Choosing the system coupling operators as $A_1=\sigma^x$ and $A_2=\sigma^y$, we obtain
$B_1^{(1)}=1/2 \sum_k \left(h_k b_k^\dagger+h_k^* b_k\right)$, 
$B_2^{(1)}=i/2 \sum_k \left(h_k b_k^\dagger-h_k^* b_k\right)$, and
$B_1^{(2)}=1/2 \sum_k \left(g_k c_k^\dagger+g_k^* c_k\right)$ with
$B_2^{(2)}=i/2 \sum_k \left(g_k c_k^\dagger-g_k^* c_k\right)$.
The Fourier transform of the bath correlation function for the first bath corresponds
to Sec.~\ref{SSspinboson}, whereas the Fourier transform matrix for the second bath is identical
to that of Sec.~\ref{SSelectronic}.
Accordingly, we obtain in the $\sigma^z$-eigenbasis
$\sigma^z \ket{a} = (-1)^a \ket{a}$ with $a \in \{0,1\}$ and
$\rho_a\equiv\bra{a}\RS\ket{a}$ 
the master equation
\bea\label{Emastermixed}
\rho_0 &=& - \left[\tilde\gamma_{1,0}^{(1)}+\gamma_{1,0}^{(2)}\right] \rho_0
+ \left[\tilde\gamma_{0,1}^{(1)}+\gamma_{0,1}^{(2)}\right] \rho_1\,,\nn
\rho_1 &=& + \left[\tilde\gamma_{1,0}^{(1)}+\gamma_{1,0}^{(2)}\right] \rho_0
- \left[\tilde\gamma_{0,1}^{(1)}+\gamma_{0,1}^{(2)}\right] \rho_1
\eea
with the dampening coefficients
$\tilde\gamma_{0,1}^{(1)}=\Gamma_1(\Omega) \left[1+\nb(\Omega)\right]$, 
$\tilde\gamma_{1,0}^{(1)}=\Gamma_1(\Omega) \nb(\Omega)$,
$\tilde\gamma_{0,1}^{(2)}=\Gamma_2(\Omega) \left[1-\fe(\Omega)\right]$, and
$\tilde\gamma_{1,0}^{(2)} = \Gamma_2(\Omega) \fe(\Omega)$,
where $\fe(\omega)$ and $\nb(\omega)$ have been defined in Eqns.~(\ref{Efermi}) and~(\ref{Ebose}), respectively, and
$\Gamma_1(\omega)=2\pi \sum_k \abs{h_k}^2 \delta(\omega-\omega_k)$ with
$\Gamma_2(\omega)=2\pi \sum_k \abs{g_k}^2 \delta(\omega-\epsilon_k)$ represent the coupling
strengths to the two baths, respectively.
The stationary state of Eq.~(\ref{Emastermixed}) is characterized by the generalized Boltzmann factor
\bea
\frac{\rho_1}{\rho_0} = \frac{\Gamma_1(\Omega) \nb(\Omega)+\Gamma_2(\Omega)  \fe(\Omega)}
{\Gamma_1(\Omega)\left[1+\nb(\Omega)\right]+\Gamma_2(\Omega)\left[1-\fe(\Omega)\right]}\,,
\eea
which is consistent with Eq.~(\ref{Estatmatrate}).


\section{Conclusions}

Under the Born, Markov, and secular approximations, quantum systems coupled to
a single bath described by inverse temperature $\beta$ and chemical potential $\mu$ relax
-- when the total Hamiltonian conserves a (quasi-)particle number --
into a stationary equilibrium ensemble that is described by the same inverse temperature and
the same chemical potential.
As long as only a single bath is involved, this also holds when the spectrum of the system Hamiltonian 
is not equidistant.

For coupling to multiple thermal baths and tridiagonal rate equations, a hypothetical non-thermal average bath is
effectively felt by the system, which will in general lead to a non-thermal stationary state.
However, when there exists only a single transition frequency as e.g.\ in two-level systems, the 
resulting stationary state may be well characterized by a single Boltzmann factor with two parameters
$\bar\beta$ and $\bar\mu$.
The possibility of creating level inversion however demonstrates that then $\bar\beta$ does not always
define an inverse temperature.

There are several interesting consequences:
Firstly, by using equilibration under side constraints one should be able to prepare
not only the ground state of a system Hamiltonian by dissipative means but also an energetically
sufficiently isolated energy eigenstate with a desired particle number when temperature and 
chemical potential of a single grand-canonical bath are tuned accordingly.
Secondly, in order to generate interesting non-equilibrium stationary states via coupling
to multiple baths, fermionic baths appear more favorable.
Beyond this, it is necessary (though not sufficient) to consider system Hamiltonians with
multiple allowed transition frequencies.
Energy-dependent tunneling rates can then significantly enhance the non-thermal signatures of the
resulting stationary state.
Finally, when the focus is on particle or thermal transport, the quantities of interest are
often the stationary current through the system and its fluctuations (noise), see e.g. 
Refs.~\cite{braun2006a,aghassi2006a}.
The calculation of these quantities heavily depends on the knowledge of the stationary state
and may therefore strongly benefit from its analytic knowledge.


\section{Acknowledgments}

Financial support by the DFG (Project \mbox{SCHA 1646/2-1}) is gratefully acknowledged.
The author has benefited from discussions with T. Brandes, M. Esposito, 
G. Kie{\ss}lich, and M. Vogl.



\end{document}